\documentclass[aps,pra,floatfix,twocolumn]{revtex4}
\newcommand{\eps}{\epsilon}
\newcommand{\dagg}{\dagger}

\begin{document}

\title{Theoretical study of lifetimes and polarizabilities in  Ba$^{+}$}
\author{E. Iskrenova-Tchoukova and M. S. Safronova}
\affiliation{Department of Physics and Astronomy, University of Delaware,
Newark, DE 19716}
\date{\today}

\begin{abstract}
The $6s-np_j$ ($n=6-9$) electric-dipole matrix elements and $6s-nd_j$ ($n=5-7$) electric-quadrupole
 matrix elements  in Ba$^{+}$ are calculated using  the relativistic all-order method. 
The resulting values are used to evaluate ground state dipole and quadrupole polarizabilities. 
In addition, the electric-dipole $6p_j-5d_{j^{\prime}}$ matrix elements and magnetic-dipole
 $5d_{5/2}-5d_{3/2}$ matrix element are calculated using the same method in order to determine the 
lifetimes of the  $6p_{1/2}$, $6p_{3/2}$, $5d_{3/2}$, and $5d_{5/2}$ levels. 
The accuracy of the $6s-5d_j$ matrix elements is investigated in detail in order to 
estimate the uncertainties in the quadrupole polarizability and $5d_j$ lifetime values. The 
lifetimes of the $5d$ states in Ba$^+$ are extremely long making precise experiments very 
difficult. Our final results for dipole and quadrupole ground state polarizabilities are 
$\alpha_{E1} = 124.15~a^3_0$ and $\alpha_{E2} = 4182(34)~a^5_0$, respectively. The resulting lifetime values 
are $\tau_{6p_{1/2}}=7.83$~ns,  $\tau_{6p_{3/2}}=6.27$~ns,  $\tau_{5d_{3/2}}=81.5(1.2)$~s, and 
  $\tau_{5d_{5/2}}=30.3(4)$~s. The extensive comparison with other theoretical and experimental values is 
  carried out for both lifetimes and polarizabilities.  
\end{abstract}
\maketitle

\section{Introduction}
The atomic properties of Ba$^{+}$ ion are of particular interest owing to the prospects of studying the
parity nonconservation (PNC) with a single trapped ion \cite{fortson:93}.
Progress on the related spectroscopy with a single Ba$^+$ ion is 
reported in \cite{koerber:02,koerber:03}, and precision measurements of light
shifts in a single trapped Ba$^+$ ion have been reported in
\cite{Ba:2005}. 
The PNC interactions  
gives rise to  non-zero amplitudes 
for transitions that are otherwise forbidden by the parity selection rules, 
such as $6s-7s$ electric-dipole transition in Cs.
The study of parity 
nonconservation in cesium \cite{wood:97,bennett:99} involving both high-precision measurements and 
several high-precision calculations provided an atomic-physics test of the standard 
model of the electroweak interactions  and  
yielded the
first measurement of the nuclear anapole moment (see \cite{Ginges:2004} for the review of 
study of fundamental symmetries with heavy atoms). 
The analysis of the Cs experiment, which
required a calculation of the nuclear spin-dependent PNC amplitude,
led to constraints on weak nucleon-nucleon coupling constants that
are inconsistent with constraints from deep inelastic scattering and
other nuclear experiments \cite{HAX:01}.
 More PNC experiments in other 
atomic systems, such as Ba$^+$, are needed to resolve this issue. 
The prospects for  measuring parity violation in Ba$^+$ have been recently discussed in 
\cite{koerber:03}. 

Ba$^+$ is also of particular interest for developing optical frequency standard \cite{Sherman:2005} and 
quantum information processing \cite{Blinov,Chen:2005} owing to the extremely
long lifetimes of $5d$ states.  
The accuracy of optical frequency standards is  
limited by the frequency shift in the clock transitions caused by 
the interaction of the ion with external fields. Therefore,
knowledge of atomic properties is needed for the analysis of the ultimate 
performance of such frequency standard. 

Another motivation for study of Ba$^+$ is an excellent opportunity for tests of theoretical 
and experimental methods, in particular in light of recent measurements of Ba$^+$ atomic
properties \cite{koerber:02,koerber:03,snow:05,Gallagher:2006,Snow:2007,Gurell:2007}. 
Ba$^+$ is a monovalent system  allowing for precise 
theoretical predictions, and, in some cases, for evaluation of the theoretical uncertainties 
that do not directly rely on the comparison with experiment. 
It is also an excellent testing case for further studies of Ra$^+$ ion, where the 
correlation corrections are expected to be larger owing to a larger core.  A project to measure PNC in
a single trapped radium ion  recently started at the Accelerator Institute (KVI) of the University of 
Groningen \cite{Ra:pnc}. 
 
 In this work, we calculate $6s-np_j$ ($n=6-9$), $6p_j-5d_{j^{\prime}}$  electric-dipole matrix elements, 
 $6s-nd_j$ ($n=5-9$) electric-quadrupole
 matrix elements, and  $5d_{5/2}-5d_{3/2}$ magnetic-dipole matrix element in Ba$^+$. 
 This set of matrix elements is needed for accurate 
 calculation of ground state dipole and quadrupole polarizabilities and lifetimes of the 
 $6p_{1/2}$, $6p_{3/2}$, $5d_{3/2}$, and $5d_{5/2}$ levels. 
 We carefully investigate the uncertainty in our values of  $6s-5d_j$ matrix elements  in order to 
estimate the uncertainties in the quadrupole polarizability and the $5d_j$ lifetime values.
It is particularly important to independently determine these uncertainties 
because of significant inconsistencies between different measurements of the 
$5d_{3/2}$ and $5d_{5/2}$ lifetimes \cite{Schneider:1979,Knab-Bernardini:1992,yu:97,Gurell:2007,
Plumelle:1980,Nagourney:86,madej:90}. There are also large discrepancies between 
experimental determinations of the 
$5d-6s$ quadrupole matrix element from the lifetime experiments and  studies of the Rydberg states of barium
\cite{snow:05,Snow:2007,Gallagher:2006}.
The experimental values of the ground state quadrupole polarizability from Refs.~\cite{Gallagher:1982,snow:05,Snow:2007}
differ by a factor of two; our value of the quadrupole polarizability is in agreement with 
Ref.~\cite{Snow:2007}. 
We note that there are no inconsistencies between  the experimental  lifetimes \cite{Andra:1976,Kuske:1978,pinnington:95}
of the $6p_j$ levels and experimental determinations of the 
electric-dipole ground state polarizability \cite{Gallagher:1982,snow:05,Snow:2007}. The experimental 
values of the electric-dipole polarizability of the Ba$^{+}$ ion in its ground state
 \cite{Gallagher:1982,snow:05,Snow:2007} are also in agreement with each other 
and our theoretical value. Our lifetimes of the $6p_{1/2}$ and $6p_{3/2}$ levels are in agreement 
with experimental values \cite{Andra:1976,Kuske:1978,pinnington:95} within expected accuracy (1\%). 

The paper is organized as follows. 
In Section~\ref{method}, we give a short description of the 
method used for the calculation of the matrix elements. 
In Section~\ref{section-e1}, we discuss the calculation of the electric-dipole polarizability and conduct 
comparative analysis of the correlation corrections to the $ns-np$ matrix elements in Ba$^{+}$, Cs, and Ca$^{+}$. 
The $6s-5d$ quadrupole matrix elements and the ground state 
quadrupole polarizability are discussed in Section~\ref{section-e2}, and the lifetimes are discussed in 
Section \ref{section-lifetimes}.  A consistency study of the $5d_j$ lifetime and ground state quadrupole polarizability
measurements in Ba$^{+}$ is presented in Section~\ref{section-e2}.

\section{Method}
\label{method}
We calculate the  reduced multipole matrix elements 
using the relativistic   all-order method \cite{blundell:89,Safronova:1999,Safronova:2007}
which is a linearized coupled-cluster method where all single and double excitations of the Dirac-Fock 
wave function are included to all orders of perturbation theory.  
The present implementation of the method is suitable for the calculation of matrix elements of any  one-body operator, i.e., 
the calculations of the E1, E2, and M1 matrix elements are carried out in the same way. 
We refer the reader to the review ~\cite{Safronova:2007} and references therein 
for the detailed description of the all-order method.  

Briefly, our starting point is the relativistic no-pair Hamiltonian \cite{brown:51}
expressed in second quantization as 
\begin{equation}
H = \sum_{i} \eps_{i} :a_{i}^{\dagg} a_{i}: 
+ \frac{1}{2} \sum_{ijkl} g_{ijkl} :a_{i}^{\dagg} a_{j}^{\dagg} a_{l} a_{k}: ,
\end{equation}
\noindent where $a_{i}^{\dagg}, a_{j}$ are  single-particle  creation and annihilation
operators, respectively, $\eps_{i}$ is the   Dirac-Fock (DF) 
energy for the state $i$, $g_{ijkl}$ are the two-body Coulomb integrals, and  :~: indicates normal order of the operators
with respect to the closed core.
The single-double (SD) all-order wave function is written as 
\begin{eqnarray} \nonumber
&& |\Psi_v^{\rm SD} \rangle  =  \left( 1 + \sum\limits_{ma}^{} \rho_{ma} a_{m}^{\dagg} a_{a} + \frac{1}{2} \sum\limits_{mna
b}^{} \rho_{mnab} a_{m}^{\dagg} a_{n}^{\dagg} a_{b} a_{a} \right. \\ 
       &&  +   \left. \sum\limits_{m \neq v} \rho_{mv} a_{m}^{\dagg} a_v 
        +  \sum\limits_{mna} \rho_{mnva} a_{m}^{\dagg} a_{n}^{\dagg} a_a a_v \right)|\Phi_v\rangle 
\end{eqnarray}
\noindent where $| \Phi_v \rangle$ is the lowest-order wave function
taken to be the frozen-core DF wave function of a state $v$.
Indices at the beginning of
the alphabet, $a$, $b$, $\cdots$,  refer to occupied core states, those in
the middle of the alphabet $m$, $n$, $\cdots$, refer to excited states,
and index $v$ designates the  valence orbital.
The all-order equations for the excitation coefficients $\rho_{ma}$, $\rho_{mv}$, $\rho_{mnab}$, and $\rho_{mnva}$ are solved iteratively with a finite basis set, and
the correlation energy is used as a convergence parameter.  
The basis set is defined in a spherical cavity on non-linear grid and
consists of single-particle basis states which are linear combinations of B-splines \cite{johnson:88}.
We use a basis set of 50 splines of order 9 in a spherical cavity of radius  80 a.u.
Such cavity size is chosen to accurately represent all orbitals of interest to the present study.  
The resulting excitation coefficients $\rho_{ma}$, $\rho_{mv}$, $\rho_{mnab}$, and $\rho_{mnva}$
are used to calculate the one-body E1, M1, and E2 matrix elements. 

The SD all-order method yielded results for the primary $ns-np_{j}$ E1 matrix elements of alkali-metal atoms
that are in agreement with experiment to 0.1\%-0.5\% \cite{Safronova:1999}.  
We note that while the all-order expression for the matrix elements contains 20 terms that are linear or quadratic
functions of the excitation coefficients, only  two terms are dominant for all matrix elements considered
in this work:
\begin{equation}
Z^{(a)} =  \sum\limits_{ma} \left( z_{am} \tilde{\rho}_{wmva} + z_{ma} \tilde{\rho}^{*}_{vmwa} \right)
\end{equation}
and
\begin{equation}
Z^{(c)} =  \sum\limits_{m} \left( z_{wm} \rho_{mv} + z_{mv} \rho^{*}_{mw} \right), 
\end{equation}
where $\tilde{\rho}_{mnab}=\rho_{mnab}-\rho_{nmab}$ and $z_{wv}$ are lowest-order matrix elements of the corresponding operator. 
In the case of the electric-quadrupole transitions studied in this work, the second term $Z^{(c)}$ is overwhelmingly 
(by an order of magnitude) larger than any other term. In such cases, it was found necessary to 
include at least partially triple excitations into the 
wave function
\begin{equation}
|\Psi_v^{\rm SDpT} \rangle   =|\Psi_v^{\rm SD} \rangle  + 
\frac{1}{6} \sum\limits_{mnrab}^{}  \rho_{mnrvab} a_{m}^{\dagg} a_{n}^{\dagg} a_{r}^{\dagg} a_b a_a a_v  | \Phi_v \rangle 
\end{equation}
and to correct single excitation coefficient  $\rho_{mv}$ equation for the effect of triple 
excitations \cite{blundell:91,Safronova:1999,Safronova:2005,safronovacs:04}. 
We have conducted such a calculation for the $6s-5d_j$, $6s-6d_j$, and $6s-7d_j$ electric-quadrupole 
matrix elements and refer to the corresponding results as SDpT values (i.e. including all single, double, and partial
triple excitations).

We note that such approach works poorly when  terms $Z^{(a)}$ and $Z^{(c)}$ are of similar
order of magnitude (such as all E1 transition considered here) owing most likely
 to cancellation of high-order corrections to terms $Z^{(a)}$ and $Z^{(c)}$. The term $Z^{(a)}$
is not directly corrected for triple excitations in the SDpT extension of the method leading to 
consistent treatment of the higher-order correlations only when the second term is overwhelmingly dominant.
 We refer the reader to Ref.~\cite{Triples:2006} for the detailed discussion of triple 
 excitations. The results of the matrix element calculation are discussed in the following sections.
 
\section{Ba$^{+}$ ground state dipole polarizability}
\label{section-e1}

The ground state dipole or quadrupole polarizability can be represented as a sum of the valence 
polarizability $\alpha_v$ and the polarizability of the ionic core $\alpha_{core}$ \cite{Safronova:1999}.
The calculation of the core polarizability assumes allowed excitations 
to any excited state including the valence shell, 
which requires the introduction of the small counter terms $\alpha_{vc}$ to subtract out $1/2$
of the contribution corresponding to the $6s$ shell excitation  \cite{Safronova:1999}. 
The core polarizabilities have been calculated in random-phase approximation (RPA) 
 in Ref.~\cite{johnson:83}. The accuracy of the RPA values is expected to be on the order of 
 5\% \cite{safronovacs:04}. We calculated the $\alpha_{vc}$ term the in the RPA approximation 
for consistency with $\alpha_{core}$ value. The valence dipole polarizability for the $6s$ state 
of Ba$^+$ is calculated as sum-over-states 
\begin{eqnarray}
\label{e1}
\alpha_{v, E1} = \frac{1}{3} \sum\limits_{n}^{}
\left( \frac{| \langle 6s||d||n p_{1/2} \rangle |^{2}}{E_{np_{1/2}} - E_{6s}} +
\frac{| \langle 6s||d||n p_{3/2} \rangle |^{2}}{E_{np_{3/2}} - E_{6s}} \right).
\end{eqnarray}
The sum over the principal quantum number $n$ in Eq.~(\ref{e1}) converges 
very rapidly and very few first terms have to be calculated to high precision.
In this work, we use SD all-order matrix elements and experimental energies for terms with 
$n=6-9$ and evaluate the remainder $\alpha_{tail}$ in the Dirac-Fock approximation. 
The contributions to the dipole polarizability are summarized in Table~\ref{tab1}. 
  We also list the absolute values of corresponding SD all-order reduced electric-dipole 
matrix elements $d$. The contribution of the terms with $n=6$ is overwhelmingly dominant. 
Therefore, the uncertainty in our calculation of the dipole polarizability 
is dominated by the uncertainties in the $6s-6p_{1/2}$ and $6s-6p_{3/2}$
matrix elements.

 To study the uncertainty in these values, we investigate the importance of the contributions from 
various
correlation correction terms and the overall size of the correlation correction. The 
contributions to the $6s-6p_{1/2}$ matrix element are summarized in Table~\ref{tab2}. The breakdown of the contributions
to the $6s-6p_{3/2}$ matrix element is essentially the same, and we do not list it here. 
We also give the breakdown of the correlation correction for the same transition in Cs and 
$4s-4p_{1/2}$ transition in Ca$^+$. Cs values are taken from Ref.~\cite{safronova:thesis}. Final
 Ca$^{+}$ value has been published in Ref.~\cite{Arora:BBR}.
 As we noted in Section~\ref{method}, only two terms
give large contributions to the correlation correction. While there are some cancellations
in the other terms, all them are at least an order of magnitude smaller.  Unfortunately, 
there is no straightforward way to evaluate the uncertainty in the $Z^{(a)}$ term (as we show in the 
later section it can be done for $Z^{(c)}$). Therefore, we can not make an uncertainty estimate that is 
independent on experimental observations. However, we note that Cs $6s-6p_{j}$ transitions are
extremely well studied by a number of different experimental approaches (see, for example, \cite{Amini:2003} 
and references therein), and all-order SD data are in agreement with Cs experimental values to 0.2\%-0.4\% \cite{Safronova:1999}.
The breakdown of terms for Ba$^{+}$ is slightly different than for Cs but is very similar to Ca$^{+}$. As expected,
 the size of the correlations is larger in Ba$^{+}$ than in Ca$^{+}$. Unfortunately, there is only one high-precision 
measurement of the $4p_j$ Ca$^{+}$ lifetimes \cite{jin} that is in significant (2\%) disagreement with high-precision 
theoretical results. Similar discrepancies existed for the alkali-metal atom measurements done 
with the same technique and 
later experiments confirmed the theory values. We refer the reader to Ref.~\cite{Arora:BBR} for more
detailed discussion of this issue. It would have been very interesting to see the $4p$ lifetimes in Ca$^+$
remeasured to resolve this issue. Based on the similar size of the correlation corrections for Cs and 
Ba$^+$, we expect similar accuracy of our data (on the order of 0.5\%). Therefore, the
 resulting accuracy of our dipole polarizability is expected to be on the order of 1\%.
We find that our value is in excellent agreement with both experimental values \cite{Snow:2007,Gallagher:1982}
 when our estimated uncertainty  is taken into account. Our results are in good 
 agreement with other theoretical calculations \cite{Lim:2004,Miadokova:1997,Patil:1997}.
We also note that the $\langle 6s|d|6p \rangle$ matrix element has been recently extracted from the $K$ splittings
of the bound $6snl$ states in Ref.~\cite{Gallagher:2006}, and the resulting value 
$\langle 6s|d|6p \rangle=4.03(12)$ is in excellent agreement with our result
$\langle 6s|d|6p \rangle=4.08$ (normalized spherical harmonics $C_{1}$ is factored out here for comparison). 

   \begin{table}
\caption{\label{tab1}  Contributions to the ground state $6s$ scalar dipole polarizability $\alpha_{E1}$ in Ba$^{+}$ in
units of $a^3_0$. Comparison with experiment and other calculations. The absolute values of corresponding SD all-order reduced electric-dipole 
matrix elements $d$ (in a.u.) are also given. }
 \begin{ruledtabular}
\begin{tabular}{lrr}
 \multicolumn{1}{c}{Contribution} &
\multicolumn{1}{c}{$d$} &
\multicolumn{1}{c}{$\alpha_{E1}$}  \\
\hline
$6s-6p_{1/2}  $&    3.3357&    40.18  \\
$6s-6p_{3/2}  $&    4.7065&    73.82  \\
$6s-7p_{1/2}  $&    0.0621&     0.06  \\
$6s-7p_{3/2}  $&    0.0868&     0.01  \\
$\alpha_{tail}$&          &     0.03  \\
$\alpha_{core}$&          &     10.61  \\
$\alpha_{vc}$  &          &     -0.51  \\
Total          &          &     124.15 \\
Expt.~\protect\cite{Snow:2007}&&  123.88(5)\\
Expt.~\protect\cite{Gallagher:1982}&& 125.5(10)\\
Theory~\protect\cite{Lim:2004}&&   123.07 \\
Theory~\protect\cite{Miadokova:1997} && 126.2 \\
Theory~\protect\cite{Patil:1997} && 124.7  
\end{tabular}
\end{ruledtabular}
\end{table} 

\begin{table}
\caption{\label{tab2} Contributions of different terms to the Ba$^+$, Ca$^+$, and Cs $ns-np_{1/2}$ reduced matrix elements in a.u.}
\begin{ruledtabular}
\begin{tabular}{lccc}
\multicolumn{1}{c}{Contribution} &
\multicolumn{1}{c}{Ba$^{+}$} &
\multicolumn{1}{c}{Cs~\cite{safronova:thesis}} &
 \multicolumn{1}{c}{Ca$^{+}$}  \\
\hline
 \multicolumn{1}{c}{} &
\multicolumn{1}{c}{$6s-6p_{1/2}$} &
\multicolumn{1}{c}{$6s-6p_{1/2}$} &
 \multicolumn{1}{c}{$4s-4p_{1/2}$}  \\
DF	      &  3.891	 & 5.278&	3.201\\
$Z^{(a)}$	& -0.387	 &-0.334&	-0.200\\
$Z^{(c)}$ & -0.209	 &-0.485&	-0.120\\
Other	    &  0.041	 & 0.019&	0.016\\
Total	     &  3.336	 & 4.478&	2.898\\
Correlation	&   16.6\%	 &  17.9\%&	10.5\%
\end{tabular}
\end{ruledtabular}
\end{table}
\begin{table*}
\caption{\label{tab3} Absolute values of electric-quadrupole $6s-5d_{3/2}$ and $6s-5d_{5/2}$  reduced matrix 
elements in Ba$^+$ calculated in different approximations in a.u. Columns labeled ``DF'' and ``III''  are lowest-order 
Dirac-Fock and third-order MBPT values, respectively. The third-order results calculated with maximum
number of partial values $l_{max}=6$ and  $l_{max}=10$ are given to illustrate the contribution of the higher partial 
waves. Breit correction is given separately. The all-order \textit{ab initio} results are given in columns labeled
 ``SD'' and ``SDpT'', respectively;
these results include contributions from higher partial waves and Breit correction.  The
corresponding scaled values are listed in columns labeled ``SD$_{sc}$'' and ``SDpT$_{sc}$''. The calculation of the 
uncertainties of the final values is described in detail in text.}
\begin{ruledtabular}
\begin{tabular}{lccccccccc}
\multicolumn{1}{c}{Transition} &
\multicolumn{1}{c}{DF}  &
\multicolumn{1}{c}{III ($l_{max}=6$)}  &
\multicolumn{1}{c}{III ($l_{max}=10$)}  &
\multicolumn{1}{c}{Breit}  &
\multicolumn{1}{c}{SD}  &
\multicolumn{1}{c}{SDpT}  &
\multicolumn{1}{c}{SD$_{sc}$}  &
\multicolumn{1}{c}{SDpT$_{sc}$} &
\multicolumn{1}{c}{Final}  \\
\hline
$6s-5d_{3/2}$ &14.76	&11.82& 11.75& -0.07& 12.42& 12.66& 12.63& 12.59& 12.63(9)\\
$6s-5d_{5/2}$ &18.38	&14.86& 14.78& -0.09& 15.55& 15.84& 15.80 &15.76& 15.80(11)
\end{tabular}
\end{ruledtabular}
\end{table*}
\section{Ba$^{+}$ ground state quadrupole polarizability}
\label{section-e2}
The valence part of the quadrupole polarizability is given in the  sum-over-states approach by
\begin{eqnarray}\label{e2}
\alpha_{v,E2} = \frac{1}{5} \sum\limits_{n}^{} 
\left( \frac{|\langle 6s||Q||n d_{3/2} \rangle |^{2}}{E_{n d_{3/2}} - E_{6s}} +
\frac{| \langle 6s||Q||n d_{5/2} \rangle |^{2}}{E_{n d_{5/2}} - E_{6s}} \right).
\end{eqnarray}
   \begin{table}
\caption{\label{tab4}  Contributions to the ground state $6s$ quadrupole polarizability $\alpha_{E2}$ in Ba$^{+}$ and their uncertainties in
units of $a^5_0$. The absolute values of corresponding  all-order reduced electric-quadrupole 
matrix elements $Q$ (in a.u.) and their uncertainties are also given.}
 \begin{ruledtabular}
\begin{tabular}{lrr}
 \multicolumn{1}{c}{Contribution} &
\multicolumn{1}{c}{$Q$} &
\multicolumn{1}{c}{$\alpha_{E2}$}  \\
\hline
$6s-5d_{3/2}  $&  12.63(9)  &1436(20) \\
$6s-6d_{3/2}  $&  16.83(5)  & 270(2)  \\
$6s-7d_{3/2}  $&   5.68(5)  & 23.7(4) \\
$6s-8d_{3/2}  $&   3.09(6)  &  6.3(3) \\
$6s-9d_{3/2}  $&   2.07(4)  &  2.7(1) \\ [0.2pc]
$6s-5d_{5/2}  $&   15.8(1)& 1932(27)   \\
$6s-6d_{5/2}  $&   20.30(6) &   392(2)   \\
$6s-7d_{5/2}  $&    6.98(6) &  35.7(6)   \\
$6s-8d_{5/2}  $&    3.83(8) &   9.6(4)   \\
$6s-9d_{5/2}  $&    2.57(5) &   4.192)   \\[0.2pc]
$\alpha_{tail}$&            &  24(6)     \\
$\alpha_{core}$&          &  46(2)  \\
Total          &          &    4182(34) \\
Expt.~\protect\cite{Snow:2007}&&  4420(250)\\
Expt.~\protect\cite{snow:05}&&  2462(361)\\
Expt.~\protect\cite{Gallagher:1982}&& 2050(100) \\
\end{tabular}
\end{ruledtabular}
\end{table} 
The RPA core value \cite{johnson:83} is $46a_0^5$, and the $\alpha_{vc}$
term is negligible. The terms containing the $6s-5d_{3/2}$ and $6s-5d_{5/2}$  
matrix elements give overwhelmingly dominant contribution to the total values. Therefore, 
we study these transitions in more detail and evaluate their uncertainties. Unlike the 
case of the E1 transitions considered earlier, 
$Z^{(c)}$ term contributes over 90\% of the total correlation correction. 
Therefore, we carried out the calculation using both SD and SDpT approaches described in Section ~\ref{method}.
We also carried out semi-empirical scaling in both approximations by multiplying single excitation 
coefficients $\rho_{mv}$ by the ratio of the ``experimental'' and corresponding (SD or SDpT)
correlation energies \cite{blundell:91}. The ``experimental'' correlation energies are determined as the 
difference of the total experimental energy and the DF lowest-order values.
 The calculation of the matrix elements is then repeated with the 
modified excitation coefficients. 
The accuracy of such scaling procedure for the similar cases was discussed in detail in 
Refs.~\cite{Safronovarb:04,safronovacs:04,Safronova:2005}. The reasoning for such a scaling procedure in 
third-order perturbation theory (scaling of the self-energy operator) has been discussed in Ref.~\cite{ADNDT:1996}.   
We list SD, SDpT, and the corresponding scaled results (labeled ``SD$_{sc}$'' and ``SDpT$_{sc}$'')
in Table~\ref{tab3}. The lowest-order DF results are listed to illustrate the size of the correlation corrections. 
We demonstrate the size of the two other corrections, contribution of the higher partial waves and 
Breit correction, in the same table. The first correction results from the truncation 
of the partial waves in all sums in all-order calculation at $l_{max}=6$. All-order 
calculation with higher number of partial waves is unpractical. Therefore, we carry out the third-order MBPT
calculation (following Ref.~\cite{ADNDT:1996}) including all partial waves up to $l_{max}=6$ and $l_{max}=10$  and 
take the difference of these two values to be the  contribution of the omitted partial waves that we add to \textit{ab initio}
all-order results. We verified that the contribution of the $l=9-10$ partial waves is very small justifying the omission of 
contributions from $l>10$.  The Breit correction is calculated as the difference of the 
third-order results with two different basis sets. The second basis set is 
generated with taking into account one-body part of the Breit  interaction. 
We note that scaled values should not be corrected for either partial wave truncation error or
Breit interaction to avoid possible double-counting of the same effects. 
We take SD$_{sc}$ values as our final results. The uncertainty of the final values is calculated as follows:
the uncertainty in the $Z^{(c)}$ term is evaluated as the spread of the most high-precision values 
(SD$_{sc}$, \textit{ab initio} SDpT, and SDpT$_{sc}$), the remaining theoretical uncertainty in the Coulomb 
correlation correction is taken to be the same as the uncertainty in the dominant $Z^{(c)}$ term.
We assume 100\% uncertainties in the contributions of the higher partial waves and Breit correction. 
The final uncertainty of the $6s-5d_j$ matrix elements (0.7\%) is obtained by adding these four uncertainties in quadrature. 
We note that this procedure for the uncertainty evaluation does not rely on the experimental values 
with the exception of the experimental energies used for scaling. 

The contributions to the ground state quadrupole polarizability are given in Table~\ref{tab4}. 
While the $n=5$ term is dominant, the contributions of the few next terms are substantial. 
Therefore, we carry out SD, SDpT, and both scaled calculations for the $6s-6d_{j}$ and $6s-7d_{j}$
matrix elements as well and repeat the uncertainty analysis described above (we omit Breit and higher-partial
wave corrections here since such precise evaluation of the uncertainties is not needed for these transitions). 
The $6s-8d_{j}$ and $6s-9d_{j}$ matrix elements are calculated in third-order MBPT, and their
accuracy is taken to be 2\% based on the comparison of the third-order and all-order values of the
$6s-7d_{j}$ matrix elements. The remainder is evaluated in the DF approximation and 
reduced by 23\% based on the comparison of the DF and third-order data for 
$6s-8d_{j}$ and $6s-9d_{j}$ matrix elements. Its accuracy is correspondingly taken to be 23\%.

Our recommended value for the ground state quadrupole polarizability  is in
  agreement within the corresponding uncertainties with the most recent 
experimental work~\cite{Snow:2007}. However, our value for the contribution of the 
$6s-5d_{j}$ transitions to the quadrupole polarizability [3368(34)] differs by about a factor of 2
from the experimental values \cite{Snow:2007,snow:05,Gallagher:2006} obtained based 
on the nonadiabatic effects on the Rydberg fine-structure
intervals. This issue and the discrepancies in the experimental values of the quadrupole 
polarizabilities are addressed in detail in 
Ref.~\cite{Snow:2007}. 
We note that these experimental values  of the $6s-5d_{j}$ contributions to the 
quadrupole polarizabilities (1524(8) \cite{Snow:2007} and 1562(93) \cite{Gallagher:2006} in the two most 
recent studies) are significantly inconsistent with all high-precision calculations of the $5d_j$
lifetimes \cite{Guet:1991,Guet:2007,dzuba:01,das:02,Sahoo:2006,Gurell:2007} carried out by different methods as well as with 
all experimental 
lifetime measurements (also carried out by different techniques)
 \cite{Schneider:1979,Knab-Bernardini:1992,yu:97,Gurell:2007,Plumelle:1980,Nagourney:86,madej:90}.
 For comparison, the value 1562(93) obtained from the $\langle 6s|r^2|5d \rangle=9.76(29)$
  matrix element that was  extracted from the $K$ splittings
of the bound $6snl$ states in Ref.~\cite{Gallagher:2006} 
corresponds to the lifetime $\tau_{5d{3/2}}=170(10)$~s 
that is a factor of 2 longer than all other values.  
  We discuss the lifetimes of the 
$5d_{3/2}$ and $5d_{5/2}$ levels in the next section.

\vspace{0.3cm}
\section{Lifetimes}
\label{section-lifetimes}

The lifetime  of a state $a$ is calculated as 
$\tau_{a} = (\sum_{b \leq a} A_{ab})^{-1}$. The E1, E2, and M1 transition rates $A_{ab}$
are given by \cite{johnson:02}:
\begin{eqnarray}
A_{ab}^{E1} &=& \frac{2.02613 \times 10^{18}}{\lambda^{3}} \frac{S_{E1}}{2j_a+1}~s^{-1},\\
A_{ab}^{E2} &=& \frac{1.11995 \times 10^{18}}{\lambda^{5}} \frac{S_{E2}}{2j_a+1}~s^{-1},\\
A_{ab}^{M1} &=& \frac{2.69735 \times 10^{13}}{\lambda^{3}} \frac{S_{M1}}{2j_a+1}~s^{-1},
\end{eqnarray}
respectively, where $\lambda$ is the wavelength of the transition in \AA~and $S$ is the line strength. 
In this work, we calculated the lifetimes of the $6p_{1/2}$, $6p_{3/2}$, $5d_{3/2}$, and $5d_{5/2}$
levels in Ba$^{+}$. The results are compared with experimental and other theoretical values in Table ~\ref{tab5}. 
Since the $6p$ levels are above $5d$ levels in Ba$^+$, we also needed to calculate 
the SD all-order reduced matrix elements for the $6p-5d$ E1 transitions, and our results (in atomic units) are 
$d(6p_{1/2}-5d_{3/2})=3.034$, $d(6p_{3/2}-5d_{3/2})=1.325$, and $d(6p_{3/2}-5d_{5/2})=4.080$. These values include 
contributions from the higher partial waves (0.6\%) and 0.1\%-0.2\% Breit correction. The correlation corrections
to these transitions are similar to the ones for the $6s-6p_j$ transitions. Therefore, similar (on the order of 0.5\%)
accuracy is expected for these matrix elements.
 The $6s-6p_j$ transitions contribute about 73\% to the respective $\sum_{b \leq a} A_{ab}$ totals for the $6p_j$ lifetimes.
Based on our evaluation of the uncertainty in these matrix elements discussed in Section~\ref{section-e1}, we expect
present $6p$ lifetime values to be accurate to about 1\%. 
Our results are in excellent agreement with other recent theoretical \cite{dzuba:01,das:02}
and experimental \cite{Andra:1976,Kuske:1978,pinnington:95} values.  
The calculation of Refs.~\cite{Guet:1991,Guet:2007} is a third-order MBPT calculation that omits 
 higher-order corrections included in the present calculation, slightly different values are expected.

   \begin{table}
\caption{\label{tab5}  Lifetimes of the $6p_j$ and $5d_j$ states in Ba$^{+}$; comparison with experiment and other theory. The lifetimes
of the $6p_j$ states are given in ns, and the lifetimes of the $5d_j$ states are given in s. }
 \begin{ruledtabular}
\begin{tabular}{lcccc}
 \multicolumn{1}{c}{}&
 \multicolumn{1}{c}{$\tau_{6p_{1/2}}$~(ns)}&
\multicolumn{1}{c}{$\tau_{6p_{3/2}}$~(ns)} &
\multicolumn{1}{c}{$\tau_{5d_{3/2}}$~(s)} &
 \multicolumn{1}{c}{$\tau_{5d_{5/2}}$~(s)} \\
\hline
Present                                  & 7.83 & 6.27 & 81.5(1.2)&  30.3(4)\\
Theory~\protect\cite{Guet:1991,Guet:2007}& 7.99 & 6.39 & 83.7     &  30.8 \\
Theory~\protect\cite{dzuba:01}           & 7.89 & 6.30 & 81.5     &  30.3 \\
Theory~\protect\cite{das:02}             & 7.92 & 6.31 & 81.4     &  36.5 \\
Theory~\protect\cite{Sahoo:2006}         &      &      & 80.1(7)  &  29.9(3) \\
Theory~\protect\cite{Gurell:2007}        &      &      & 82.0     &  31.6  \\
Expt.~\protect\cite{Andra:1976}          &      & 6.312(16)  & & \\
Expt.~\protect\cite{Kuske:1978}          & 7.92(8) &  & & \\
Expt.~\protect\cite{pinnington:95}       &7.90(10)&6.32(10)&      &    \\
Expt.~\protect\cite{Schneider:1979}       &      &      & 17.5(4) &         \\
Expt.~\protect\cite{Knab-Bernardini:1992} &      &      & 48(6) &         \\
Expt.~\protect\cite{yu:97}               &       &      & 79.8(4.6)&    \\
Expt.~\protect\cite{Gurell:2007}         &       &      & 89.4(15.6)&  32.0(4.6)  \\                      
Expt.~\protect\cite{Plumelle:1980}        &      &      &       & 47.0(16)   \\
Expt.~\protect\cite{Nagourney:86}       &      &      &       & 32.0(5)   \\
Expt.~\protect\cite{madej:90}            &       &      &          & 34.5(3.5)    \\
\end{tabular}
\end{ruledtabular}
\end{table} 
Only one transition contributes to the $5d_{3/2}$ lifetime: $6s-5p_{3/2}$ E2 transition (the 
contribution of the $6s-5d_{3/2}$ M1 transition is negligible). In the case of the $5d_{5/2}$ lifetime,
M1 $5d_{5/2}-5d_{3/2}$ transition has to be included as pointed out in \cite{dzuba:01,Guet:2007,Sahoo:2006}.
Our SD all-order value for this transition (in a.u.)  is 1.5493. The correlation correction 
contribution is very small, and the lowest order gives essentially the same value, 1.5489.
The M1 transition contributes 17\% to the   $\sum_{b \leq a} A_{ab}$ total for the $5d_{5/2}$ level.

We compare our final results for the $5d_{3/2}$ and $5d_{5/2}$ lifetimes with experimental \cite{Schneider:1979,Knab-Bernardini:1992,yu:97,Gurell:2007,Plumelle:1980,Nagourney:86,madej:90}
and other theoretical \cite{Guet:1991,Guet:2007,dzuba:01,das:02,Sahoo:2006,Gurell:2007}
 values  in Table~\ref{tab5}. We note that calculation \cite{das:02} omitted $5d_{5/2}-5d_{3/2}$ M1 contribution to the $5d_{5/2}$
lifetime leading to a higher value, as noted in later work \cite{Sahoo:2006}.
Our results are in agreement with other theoretical calculations, most recent values
 from \cite{Gurell:2007} measured in a beam-laser experiment performed at the ion storage ring CRYRING, as
 well as experimental values from \cite{yu:97,Nagourney:86,madej:90}.

\section{Conclusion}

In conclusion, we carried out the relativistic all-order calculations of 
 Ba$^+$ $6s-np_{j}$ ($n=6-9$), $6p_{1/2}-5d_{3/2}$, $6p_{3/2}-5d_{5/2}$, 
  and $6p_{3/2}-5d_{5/2}$ electric-dipole  matrix elements; $6s-5d_{3/2}$, $6s-5d_{5/2}$, 
  $6s-6d_{3/2}$, $6s-6d_{5/2}$, $6s-7d_{3/2}$, and $6s-7d_{5/2}$
   electric-quadrupole matrix elements; and $5d_{5/2}-5d_{3/2}$ magnetic-dipole matrix element. 
   These values are used to evaluate lifetimes of the $6p_{1/2}$, $6p_{3/2}$, $5d_{3/2}$, and $5d_{5/2}$
   levels as well as dipole and quadrupole ground state polarizabilities.
   Extensive comparison with other theoretical and experimental values is carried out.  
  The present values of the dipole polarizability and $6p_{j}$ lifetimes  are in excellent agreement with experimental values.
  We estimated the uncertainty of our theoretical values for these properties to be on the order of 1\%. 
   Our recommended value of the quadrupole ground state polarizability $\alpha_{E2} = 4182(34)a^5_0 $ is 
     in agreement with the most recent experimental work \cite{Snow:2007}.
      Our recommended values for the $5d_j$ lifetimes $\tau_{5d_{3/2}}=81.5(1.2)$~s and 
  $\tau_{5d_{5/2}}=30.3(4)$~s  are in agreement with other theoretical calculations, most recent values
 from \cite{Gurell:2007} measured in a beam-laser experiment performed at the ion storage ring CRYRING, as
 well as experimental values from \cite{yu:97,Nagourney:86,madej:90}.

\section*{Acknowledgements}
We thank Steve Lundeen and Erica Snow for many useful discussions.
This work was supported  in part by the National Science Foundation Grant 
No.\ PHY-04-57078.

\end{document}